\def \cm{~\rm{cm}}
\def \s{~\rm{s}}
\def \km{~\rm{km}}

\def \K{~\rm{K}}
\def \g{~\rm{g}}

\def \AU{~\rm{AU}}

\def \yr{~\rm{yr}}

\documentclass[onecolumn]{aa}
\usepackage{graphicx}

\usepackage[german,english]{babel}
\begin{document}

%

\title{Interaction of young stellar object jets with their accretion disk
\thanks{ Research supported by the Israel
Science Foundation}}

\author{Noam Soker\inst{1}}

\offprints{Noam Soker, \email{soker@physics.technion.ac.il}}
\institute{Department of Physics, Technion-Israel Institute of
Technology,  32000 Haifa, Israel; soker@physics.technion.ac.il }
\date{Received ---- / Accepted ----}

\titlerunning{Jet-disk interaction}

\abstract{I reexamine recent observations of velocity profiles
across jets blown by young stellar objects, and argue that the
observations do not support the interpretation of jets rotating
around their symmetry axes. Instead, I propose that the
interaction of the jets with a twisted-tilted (wrapped) accretion
disk can form the observed asymmetry in the jets' line of sight
velocity profiles. The proposed scenario is based on two plausible
assumptions.
(1) There is an inclination between the jet and the
outer parts of the disk; the jet is perpendicular to the
inner part of the disk;
Namely, there is a twisted-tilted (wrapped) disk.
(2) The disk-jet interaction slows down the jet as the jet
entrains mass from the disk, with larger decelaration of jet
segments closer to the tilted disk.
The proposed scenario can account for the basic properties of the
observed velocity profiles, while having the advantage that there
is no need to refer to any magnetic jet launching model, and there
is no need to invoke jet rotation with a huge amount of angular
momentum. } \maketitle


\section{Introduction} \label{sec:intro}

Asymmetry in the line of sight velocity across some jets ejected by
young stellar objects (YSOs) have been interpreted as being due
to a large scale rotation of the material in each jet around the jet's
symmetry axis (e.g., Bacciotti et al.\ 2002 and Coffey et al.\ 2004,
hereafter B2002 and C2004).
A rotation around the jet axis is predicted by the
magneto-centrifugal models for jet launching.
In these models the magnetic fields that are anchored into the
accretion disk-star system play a dominate role in accelerating
the jet's material from the accretion disk.
Most models in YSOs are based on the operation of large scale magnetic
fields driving the flow from the disk; either via the ``centrifugal wind''
mechanism, first proposed by Blandford \& Payne (1982), or from a narrow
region in the magnetopause of the stellar field via an ``X-wind
mechanism'' introduced by Shu et al. (1988, 1991) and in a
somewhat different setting, by Ferreira \& Pelletier (1993, 1995).
See recent reviews and papers by K\"onigl and Pudritz (2000), Shu et al.
(2000), Ferreira (2002), and Krasnopolsky et al. (2003).
It should, however, be pointed out that the origin of the large scale
magnetic fields and the manner that open field lines of sufficiently
strong magnitude persist (in the centrifugal wind models), or the
manner by which a stellar field interacts with the disk, allowing
inflow and at the same time driving an outflow (in the X-wind models)
are still open key issues of the theory (e.g. Heyvaerts et al.\ 1996).
In addition, it seems that thermal pressure is needed in
some of these models (e.g., Ferreira \& Casse 2004).

In a recent paper Cerqueira \& de Gouveia Dal Pino (2004)
conduct magnetohydrodynamics (MHD) 3D numerical simulations
of YSO jets with rotation.
They use parameters suitable for the jet of DG Tauri, a YSO
studied by B2002, and obtain radial velocity maps which are in
good agreement with the observations of B2002.
In the appendix I point out some difficulties with this
interpretation of jet rotation.

Magnetic fields might also play a role in triggering the jet's ejection
event, e.g., by causing instabilities in the disk, and
in collimating the jets (e.g., Heyvaerts \& Norman 1989).
The collimation issue is, however, still quite controversial and while
magnetic collimation is certainly plausible, its exact nature is
probably quite involved and still not fully understood (see the
recent works of Bogovalov \& Tsinganos 2001 and Li 2002).
In some models external pressure are required to maintain collimation,
e.g., Kato et al. (2004), and in the magnetic tower model
of Lynden-Bell (2003).
Although some of the arguments in these papers apply to
relativistic jets, I mention them here as I am looking for a unified
collimation machanism in different jets environments.
In section (2) I discuss some aspects of disk-jet interaction
in YSOs.

In some models, magnetic fields do not play a major role
in launching the jet.
These include the thermally-launched jet model (Torbett 1984;
Torbett \& Gilden 1992), that was applied to YSOs
(Soker \& Regev 2003) and cataclysmic variables
(Soker \& Lasota 2004).
Motivated by non-magnetic models, in section (3) I show that the
asymmetric radial velocity field of these YSO jets may be
interpreted in an alternative way, where the magnetic field
have no major role.
In this proposed model, instead of jet rotation, the velocity field is
attributed to the interaction of the jet with the accretion disk wind
and/or corona.
A summary of the main results is in section (4).

\section{Interaction of a fast outflow with the disk}
\label{sec:disk}

I here discuss an alternative explanation to the results of
C2004, based on the interaction of an outflow (jets) blown
at large speeds from the vicinity of the central star
with the accretion disk, mainly via entrainment of disk's
mass by the jets.
The entrainment process itself was studied in the past.
The entrainment is likely to occur in a turbulent mixing layer
between the jet and the ambient gas (e.g., Kahn 1080;
Canto \& Raga 1991).
While most studies consider entrainment of (molecular) gas
at distances of hundreds of AU from the source (e.g. Davis et al. 2004),
I consider entrainment of disk material at a distance of
several AU from the source.
Based on CO observations, Stahler (1994) build an empirical model for
entrainment in YSO jets on scales of hundreds AU.
According to this model, the velocity of matter
inside the jet decreases with increasing distance from the jet's axis.
This velocity profile is assumed here as well.
The next step in developing the presently proposed scenario, will be
to include such an effect for an interaction of a jet with a wrapped
disk, but on a scale of several AU rather than hundreds of AU. .

First I note that accretion disks around YSOs can be quite extended in
the vertical direction (e.g., Semenov et al. 2004; Malbet et al. 2001).
In addition, the disk may blow a slow wind from an extended region,
e.g., due to coronal type magnetic activity on the surface of the disk.
I assume therefore that a fast wind, hereafter refer to as jets,
blown from the vicinity of the central star at a speed of
$v_j \sim 300-500 \km \s^{-1}$, interacts with the disk and/or
its wind.
The outer parts of the jets are influenced the most.
Let us first assume that the outskirts of the jet are strongly shocked
at a distance of $r_s$ from the central star.
The density of the post-shocked gas, which is $\sim 4$ times the
pre-shocked density, is given by
\begin{equation}
\rho_s \simeq 3 \times 10^{-19}
\frac{\dot M_j}{10^{-8} M_\odot \yr^{-1}}
\left( \frac{r_s}{10 \AU} \right)^{-2}
\left( \frac{v_j}{300 \km \s^{-1}} \right)^{-1} \g \cm^{-1}
\label{dens1}
\end{equation}
where $\dot M_j$ is the mass loss rate into the two jets together.
I assumed above a spherically symmetric fast wind (jet);
an initial (poor) collimation means a higher jet's density.
On the other hand, higher mass loss rate near the jet axis
(e.g., Liseau et al. 2005)
means a lower density on the jet's outskirts.
The density given in equation ({\ref{dens1}}) is lower than
the disk's surface density
$\rho_{d} \simeq 10^{-17} \g \cm^{-1}$ at $r_d=10 \AU$ in the
model of Semenov et al.\ (2004) for an accretion rate of
$\dot M_{\rm acc} = 10^{-7} M_\odot \yr^{-1}$.
The gas shocked to a temperature of
\begin{equation}
T_s = \frac{3}{16}\frac{\mu m_H}{k} v_j^2
= 1.3 \times 10^6
\left( \frac{v_j}{300 \km \s^{-1}} \right)^{2} \K.
\label{shockt}
\end{equation}
For this temperature I approximate the cooling curve
(e.g., Gaetz, Edgar, \& Chevalier 1988) by
$\Lambda = 3.5 \times 10^{-23} (T/10^7~{\rm K})^{-1/2}
~{\rm erg}~{\rm cm}^3~{\rm s}^{-1}$, for
$2 \times 10^5 \la T \la 2 \times 10^7 \K$.
The cooling rate, power per unit volume, is given by
$\Lambda n_e n_p$, where $n_e$ and $n_p$ are the
electron and proton density, respectively.
I take the cooling time at a constant pressure
$\tau_c = (5/2) n k T/\Lambda n_ e n_p$ for a fully ionized post-shock gas.
For the flow time I take $\tau_f = r_s/v_j$, which is about the time
required for the high post shock pressure to
re-accelerate the post shock gas.
Relating the shock temperature to the speed of a fully ionized solar
composition wind, and taking the density from equation ({\ref{dens1}}),
gives
\begin{equation}
\chi \equiv \frac{\tau_c}{\tau_f} \simeq 10
\left( \frac{ \dot M_j }{10^{-8}~M_\odot~{\rm yr}^{-1}} \right)^{-1}
\left( \frac{ r_s }{10~{\rm AU}} \right)
\left( \frac{ v_j}{300~{\rm km}~{\rm s}^{-1}} \right)^5 .
\label{eq:chi1}
\end{equation}
The reacceleration distance is shorter than $r_s$, and the reacceleration
time shorter than $r_s/v_j$.
Also, the shock is oblique, and less wind's kinetic energy is thermalized.
Therefore, the value given in equation ({\ref{eq:chi1}}) is a lower limit.

From equations ({\ref{dens1}})-({\ref{eq:chi1}}) the following conclusions can be drawn.
\newline
(1) The density in the surface layer of disks around YSOs can
be higher than that in the jet, hence influence its propagation.
The main modes of interaction are the entrainment of disk's mass
by the jets, a process that slows down the region of the jets
entraining the mass, and the deflection of the jet's outskirts
via a strong shock, a process which recollimate the jet, and which
may deflect somewhat the jet.
\newline
(2) The shocked material is heated to soft X-ray emission temperatures,
and can cool on a time scale longer than the time it is
being re-accelerated.
Namely, there is an adiabatic-type flow.
Note that the fraction of the mass in the jet that is
shocked to X-ray emitting temperature is very low, as only the
outskirts of the jets are shocked.
Pressure gradient then collimate the inner regions of the jet,
and entrainment slows the jet down.
Therefore, the X-ray emission from the shocked region will be weak.
This is compatible with the {\it Chadra} observations of
Getman et al. (2002), who find that the presence or absence of an outflow
does not appear to produce any difference in the X-ray properties
of YSOs.
\newline
(3) The relatively small energy loss implies that the re-acclerated
gas reaches velocity close to its original speed.
\newline
(4) However, because of the high disk-density in the interaction
region, even a small-volume mixing implies that substantial disk
material is entrained by the jet outskirts.
This, and some radiative cooling, will slow down the re-accelerated gas.
\newline
(5) The last point implies that if large amount of gas from the
disk is entrained by the fast jet, a slow dense jet might
be formed.
\newline
(6) The inner boundary of the surface layer in the disk
model of Semenov et al.\ (2004) mentioned above is at a height
of $z_U=0.17 (r_d/\AU)^{1.27} \AU$.
The outer boundary of the surface layer flares even faster.
If the surface layer redirect the outskirts of the jets, this
flaring will lead to a collimation of the jet.
\newline
(7) The cooling time is not much longer than the flow time.
Because of inhomogeneities, some blobs may still cool.
I attribute the differences in the velocities observed in
different lines by C2004, to such inhomogeneities, which
possibly led to thermal instabilities as well.

\section{Interaction with the disk and radial velocity profiles}
\label{sec:depart}

In the previous section I discussed the interaction of
a fast wind (2 oppositely ejected jets) with the accretion disk.
This might result in jets' collimation, where the basic collimating
force is thermal pressure, with possible help from a ram
pressure of a slow disk's wind.
Magnetic hoop-stress might still play a role, but not
necessarily a dominate one.
In this section I show that in principle the radial-velocity gradient
in the jets observed by B2002 and C2004 can be explained by the interaction
of the jets with the accretion disk, if one assumes that the
accretion disk is tilted with respect to the jet's axis.
What C2004 attribute to rotation, I attribute to different
degree of slowed down jet's segment because of a tilted
accretion disk.
Since the jets will be blown perpendicular to the inner
disk, what is required is that the inner disk will be twisted with
respect to the regions of the disk further out.
These are the outer parts of the disk that the jets interact with.
Such a twisted (wrapped) disk can result from radiation induced warping
(Armitage \& Pringle 1997; see figs 7 and 10 of  Wijers \& Pringle
1999 for the geometry of a twisted disk), from
the influence a stellar binary companion (Larwood et al. 1996),
or possibly from the influence of a massive planet.

The assumed disk structure is schematically drawn in Figure 1.
The $z$ axis is taken along the jet axis, with the positive side
in the receding (red-shifted) jet direction.
The $xz$ plane is taken to be the plane defined by the jet axis and
the line of sight, with the positive $x$ axis on the
observer side.
The angle between the approaching (blue-shifted) jet's axis and the
line of sight is $\beta$.
The observed section of the jet is assumed to be cylindrical, with
radius $R_j$.
The $y$ axis is perpendicular to the jet's axis and to the line of sight
(Fig. 2).
The first basic assumption is that the disk plane is inclined to the $xy$
plane: it has a maximum height toward the $+z$ direction at an angle
$\phi_0$ from the $x$ axis, and a maximum height in the $-z$ direction
at an angle of $\pi+\phi_0$ relative to the $x$ axis.
The tilt angle is $\alpha$.
The second basic assumption is that the disk slows down the jet,
with larger decelaration of jet segments closer to the tilted disk,
and less efficient deceleration closer to the jets' axis and away
from the disk.
I also consider the regions of jets far from the origin, after the
different segments have reached their terminal speed,
namely, $v_j(r,\phi)$ does not depend on $z$.
I parameterize the slowing down of the jet material by taking the
velocity at a distance $r$ from the jet axis and at an angle
$\phi$ to the $x$ axis as
\begin{equation}
\frac {{\vec{v}}_j(r,\phi)}{v_{ja}} =  k \hat z -
C_1 [ k +f_2(\alpha) \cos (\phi-\phi_0) ]
\left( \frac{r}{R_j} \right)^n \hat z,
\label{vjet1}
\end{equation}
where $C_1>0$, which is a constant, and $1> f_2(\alpha) \ge 0$, which
is a monotonic function of the tilt angle $\alpha$, depends also on
the strength of the jet-disk interaction, $v_{ja}$ is the magnitude of
the outflow speed along the jets' axis, and
\begin{eqnarray}
 k \equiv \left\{
\begin{array}{cl}
 1 &  {\rm for \ receding \ jet }  \\
 -1 &  {\rm for \ approaching \ jet }.
\end{array}
 \right .
\label{kval}
\end{eqnarray}

The observation at specific point across the jet includes all
jets' segments along the line of sight to that point.
To get a sense to the range of the observed velocities, I draw in
figure 3 the component of the jet's radial velocity along the
line of sight, as a function of the distance from the jet's axis,
for three locations along the line of sight for each point:
\newline
($i$) The close side (to the observer) of the jet along
the line of sight (marked $i$ in figures 1 and 2; thick solid line in figure 3).
In this case the point is located at $(r,\sin \phi)=(R_j,y/R_j)$, with
$\cos \phi = +\sqrt{1-y^2/R_j^2}$.
Substituting this value in equation ({\ref{vjet1}}) yields for the
jet velocity component along the line of sight
\begin{equation}
\frac {v_j}{v_{ja}} =  \left[ k -
C_1 k - C_1 f_2 \left( \sqrt{1-\frac{y^2}{R_j^2}} \cos \phi_0
+\frac{y}{R_j} \sin \phi_0 \right) \right] \cos \beta .
\label{vjeti}
\end{equation}
($ii$) The far side edge of the jet (marked $ii$ in figures
1 and 2;
thick dashed line in figure 3).
In this case the point is located at $(r,\sin \phi)=(R_j,y/R_j)$,
with $\cos \phi = -\sqrt{1-y^2/R_j^2}$.
Substituting this value in equation ({\ref{vjet1}}) yields for the
jet velocity component along the line of sight expression like that in equation
({\ref{vjeti}}), but with a minus sign in front of the square root.
\newline
($iii$) The point closest to the jet's axis, namely, on a line through
the jet's axis and perpendicular to the line of sight
(marked $iii$ in figures 1 and 2; thin solid line in figure 3).
This gas element resides at distance $y$ from the jet axis, and
perpendicular to the line of sight (i.e., in the $yz$ plane).
Substituting in equation ({\ref{vjet1}}) $r=\vert y \vert$ (see Fig. 2),
and $\theta= \pi/2$ and $\theta =3 \pi /2$, for $y>0$ and $y<0$, respectively,
gives
\begin{equation}
\frac{{\vec{v}}_j(r,\phi)}{v_{ja}} = k \cos \beta -
C_1 \left[ k +f_2 \frac{y}{\vert y \vert} \sin \phi_0  \right]
\left( \frac{\vert y \vert }{R_j} \right)^n \cos \beta.
\label{vjetiii}
\end{equation}

The following conclusions follow from figure 3, regarding the proposed scenario
for jet-disk interaction.
\begin{enumerate}
\item This scenario explains the decline in radial velocity with
increasing distance from the jet's axis. Such a behavior is clearly
observed in DG Tauri (fig. 1 of B2002), the red lobe of RW Aurigae
(fig 5. of C2004), and in LkH$\alpha$ 321 (fig 5. of C2004).
\item An inclination of the jet to the outer disk with which it
interacts (the inner disk, $r_d \la 0.1-1 \AU$, which blows the jet
is perpendicular to the jet), can explain the velocity shift between the
two sides of each jet, with the same sense of shift in the blue shifted
(approaching) and red shifted (receding) jets.
\item Different regions along each line of sight have different
projected velocity.
The variation increases toward the jet's symmetry axis.
If different spectral lines originate in different regions in the jet,
then there will be large variations in the projected velocities
between different spectral lines.
This might be the case, for example, if instabilities in the jet-disk
interaction region lead to the formation of dense clumps, where the
emission of some spectral lines are favored over other spectral lines.
The instabilities might occur in some places, but not in others.
Large differences in velocity profiles between different spectral lines
are observed in a large fraction of the jets, e.g.,
in the red lobe of TH28 (fig 5 of C2004).
The data are too noisy to tell whether the variation near the jet's center
is larger.
\item Large velocity differences between the two jet's sides can be
explained without invoking a huge amount of angular momentum in the jet.
\end{enumerate}

\section{Summary} \label{sec:sum}

The goal of the present paper is to provide an interpretation
to the velocity profiles of some YSO jets without referring to
any magnetic jet launching model.
These observations (B2002; C2004) show that the two sides of some YSO
jets have different projected velocity (line of sight velocity), and
that the sense of the velocity shift is the same in the blue
(approaching) and red (receding) jets.
B2002 and C2004 interpret the velocity profiles as rotation of the jets
around their symmetry axis.
Such a rotation is expected in some magnetic jet launching
models (Cerqueira \& de Gouveia Dal Pino 2004).

In the appendix I show that the required angular momentum for the
rotation explanation is on the upper limit of what the accretion
disk can supply.
In other words, to account for the required angular momentum and the
kinetic energy of the jets, the mass outflow must be smaller than
observed.

Alternatively, I propose a scenario where the velocity profile results
from the interaction of the jets with the accretion disk.
In section (2) I argue that such an interaction is likely
to occur, as accretion disks around YSOs might have a
thick surface layer.
I then (section 3) presented a phenomenological scenario based on two
assumptions.
(1) There is an inclination between the jet and the disk further out.
Namely, the inner disk is perpendicular to the jet, but then the disk
flares in a point-symmetric manner: one side up and the opposite
side (relative to the central star) down (figure 2).
(2) The disk slows down the jet, with larger decelaration of jet
segments closer to the tilted disk, and less efficient deceleration
closer to the jets' axis.
The velocity of the gas inside the jet as function of distance from the
jet's symmetry axis $r$, and the angle relative to the direction
where the disk flares up (in the $+z$ direction) $\phi-\phi_0$,
was parameterized in equation ({\ref{vjet1}}).
The different values appearing in that equation are defined in
figures 1 and 2.
The calculated projection velocity profiles for 4 cases
are plotted in figure 3.

The conclusions driven from figure 3 are summarized in the last
paragraph of section 3.
In short, these are:
(1) The proposed non-magnetic scenario explains the decline in
radial velocity with increasing distance from the jet's axis.
(2) The inclined jet-disk scenario might explain the velocity shift
between the two sides of each jet, with the same sense of shift
in the blue shifted (approaching) and red shifted (receding) jets.
(3) The large velocity variations between different spectral lines
can be accounted for if different lines originate in different
regions in the jets.
For example, instabilities in the jet-disk interaction leads to
the formation of dense clumps where some lines are favored over
other lines.
(4) There is no need to invoke jet rotation with
a huge amount of angular momentum.

The interaction of the warrped-tilted disk with the jet will
cause a point-symmetric structure.
High spatial resolution imaging may reveal such a structure
close to the star.
In Th 28, the H$\alpha$ image of the two HH objects
possesses a point symemtric structure (Krauter 1986),
which most likely comes from precessing jets.
The two jet in RW Aur are highly symemtric, and show no
point-symmetry  (Dougados et al.\ 2000).
The one jet of DG Tau as presented in the high veloicty (HV)
map by Lavalley-Fouquet et al.\ (2000; the upper row of their
fig. 1) possesses some wiggling.
In any case, the predicted point-symmetry morphology is
of small degree, but still might be noticed.

Another prediction of the proposed scenario is that in some cases,
where disk changes twisting (wrapping) direction in a relatively short time
period, the sense of asymmetry will change.
Namely, the side which presently has a lower radial velocity,
might slowly, say on a time scale of $\sim 100 \yr$, change
to have a higher radial velocity.
However, in most disks the expected time scale is very long,
according to the different mechanisms that cause such twisting
(or titling or warping;
Armitage \& Pringle 1997; Larwood et al. 1996).

{}

{\bf APPENDIX 1: Differential radial velocity along YSO jets (will
appear in electronic version of A\&A)}
\newline

In a recent paper for three YSOs, C2004 present observed profiles
of line-of-sight velocity as function of the distance from the
jet's symmetry axis. They interpret their observations as
indicating jet rotation, and claim that the magnetocentrifugal jet
launching model is consistent with their results.
This is problematic on both accounts.
More sophisticated and detailed MHD models for jet rotation,
though, might overcome these problems.
By examining the radial velocity
figures of C2004, I note the following issues.
\newline
(1) {\bf Stochastic velocity profiles.}
Different lines show very different profiles, as is most
evidence from their figure 5.
In the red lobe of TH28, for example, the difference in velocity
between the [NII]$\lambda$6548 and the [SII]$\lambda$6716 is
$\sim 30 \km \s^{-1}$ in the SW side, while the difference is
almost an order of magnitude smaller in the NE side of the
jet axis in the same lobe.
In the YSO LkH$\alpha$ 321 the difference between these two lines
in the SE side reaches a value of $\sim 40 \km \s^{-1}$.
These differences are larger than the claimed rotational velocity
of the corresponding jets.
Later, I attribute these differences between the velocities observed in
different lines, and the difference between the two sides of the jets'
axes, to the interaction of the jet with a tilted disk (section 3).
\newline
(2) {\bf Angular momentum.}
If the radial velocity gradient is attributed to the jet's rotation,
then the specific angular momentum of the gas at the edge of
the jets must be extremely high.
For example, in the redshifted lobe of RW Aur the toroidal velocities are
$v_\phi=7-17 \km \s^{-1}$ (see last paragraph of section 3 in C2004),
in the range $y = 14-28 \AU$, where $y$ is the
distance from the jet's axis.
The specific angular momentum is in the range
$j \sim 100 - 500 \AU \km \s^{-1}$.
For an accretion disk around a central object of mass $1 M_\odot$,
this range corresponds to disk radii in the range
$r_d \sim 10-250 \AU$.
In the magnetocentrifugal jet launching model the magnetic stress
transports energy and angular momentum from the accreted mass to
the ejected mass.
In RW Aur the ejected mass is $\ga 5 \%$ the accreted mass
(Woitas et al.\ 2002), namely
$\dot M_{\rm jet} \ga 0.05 \dot M_{\rm acc}$,
and C2004 find the foot point
(where the streamline leaves the disk) of the outermost
stream line to be at $r_{fp} =1.3 \AU$.
The inner-most foot point is at $r_{fp}=0.4 \AU$.
The specific angular momentum of Keplerian disk material
at $r_d=1.3 \AU$ is $\sim 0.07$ times the specific angular
momentum of matter flowing on the outer stream line.
This implies a very efficient angular momentum transfer.
The accreted mass required to give all of its angular momentum
is $\sim 14$ times the mass ejected in the outer parts of the jet
(for $\dot M_{\rm jet} = 0.05 \dot M_{\rm acc}$).
For the average of the jet's specific angular momentum I take
$j_a \simeq 300 \AU \km \s^{-1}$, which is also the
value used in the numerical simulations of
Cerqueira \& de Gouveia Dal Pino (2004).
The total rate at which the jet carries angular momentum is then
$\dot J_j= \dot M_{\rm jet} j_d = 15 \dot M_{\rm acc} \AU \km \s^{-1}$.
In an undisturbed disk, the gas losses angular momentum
as it flows from $r_d=1.3 \AU$ to $r_d = 0.4 \AU$
(from outer to inner foot point) at a rate of
$\dot J_d= 15 \dot M_{\rm acc} \AU \km \s^{-1}$.
This implies that if C2004 interpretation of the
jets in RW Aur is correct, then according to the jet-disk models
they use, the disk must lose {\it all} of its angular
momentum already at $r_d=0.4 \AU$.
This is a much larger radius than that of the accreting
central star, which implies that the disk is truncated at
a very large redius, where escape speed is much below the
jet's speed.
\newline
(3) {\bf Energy.}
In the analysis of Anderson et al.\ (2003; see also 2005), the poloidal
speeds along jets' stream lines of the YSO DG Tauri are $<3$ times what would be
the escape velocity at the foot points of the stream lines.
In the numerical calculations of Krasnopolsky et al.\ (2003) the
poloidal terminal speeds along jet's streamlines are $\la 2$
times the escape velocity at the foot point of the stream lines.
The velocities found by C2004 (their table 4) are much higher than
the theoretical expectation; along most stream lines the terminal
jets' speeds found by C2004 are $\ga 3$ time the escape speeds.
In the YSO LkH$\alpha$ 321, for example, the jet speed is
$\sim 4-4.5$ times the escape velocity.
The required efficiency of energy transfer from accreted to ejected
mass seems to be higher than that expected in the magneto-centrifugal
models.
Higher velocities are obtained in the numerical simulations
of Garcia et al.\ (2001) which includes the thermal state of the jet.
The major heating process of the jet is ambipolar diffusion.
High poloidal velocities are obtained along stream lines,
$\sim 10$ times the escape velocity from the footpoint of
the stream line.
However, the fraction of ejected mass (out of the accreted
mass) is $1-2 \%$, much below the value of $\sim 10 \%$ in
RW Tau (Woitas et al.\ 2005).
But, again, in this section I am dealing mainly with C2004
interpretation of their observations, and not with the
general MHD model.
\newline
(4) {\bf Model versus observations.}
C2004 use the jet launching model as presented by Anderson et al.\
(2003), who use it for the YSO DG Tau.
However, in that model the toroidal velocity decreases with distance
from the jet's axis, while in the observations, both of DG Tau
(Anderson et al.\ 2003) and the three YSOs studied by C2004, the
toroidal velocity increases with distance from the jet's axis.
The opposite behavior is attributed by C2004 to projection
and beam smearing effects.
Still, despite these effects they compare their results with
the theoretical model.
I find this unjustified.
In a recent paper Pesenti et al. (2004) present a more detailed
MHD model based on jet rotation to account for the observation
of DG Tau (B2002).
Pesenti et al. (2004) find a good fit between their model and the
velocity map of DG Tau.
However, the model of Pesenti et al. (2004) for
DG Tau has two significant differences from that of TW Aur
(Woitas et al.\ 2005).
First, their derive typical toroidal velocity for stream lines
with footpoints of $\sim 1 \AU$ is much lower than that in TW Aur.
In TW Aur the central star mass is $M_\ast=1.3 M_\odot$,
and in DG Tau model of Pesenti et al. (2005) it is
$M_\ast=0.5 M_\odot$.
Velocities should be scaled like the square root of the
central mass ratio $(1.3/0.5)^{1/2}=1.6$.
Namely, the specific angular momentum in the outer jet radii in the
model of Pesenti et al. (2004), of $\sim 80 \AU \km \s^{-1}$,
is much lower than that the value of $\sim 300-500 \AU \km \s^{-1}$
found by Woitas et al.\ (2005) for the jets in TW Aur.
The second difference involved the theoretical model used by
Pesenti et al. (2004).
Pesenti et al.\ (2004) find  that only the warm jet model fit the
observations.
In this model the jet is thermally-driven (Casse \& Ferreira 2002).
Both these differences, that the jets have low specific angular
momentum, and are thermally driven, are along the main
theme of the present paper, although significant differences exist
between the warm MHD model used by Pesenti et al.\ (2004) and
the model proposed here.

{\bf APPENDIX 2: Will not appear in the Journal at all}
\newline
In many magneto-centrifugal models it is assumed that the
outflowing jet material follows the magentic field lines, which
are anchored to the disk in a steady state flow structure.
However, the jet carries with it energy and angular momentum, and
if mass flow rate is high, this will alter the disk structure. To
see this, I elaborate on the calculation presented in point (2) of
section 2. Let us take the specific angular momentum and specific
kinetic energy of the gas in the jet to be $j_j=300 \AU \km
\s^{-1}$ (I neglect the angular momentum carried by the magnetic
field), and $e_j=v_j^2/2=4.5\times 10^4 \km^2 \s^{-2}$,
respectively, as in the numerical simulations of Cerqueira \& de
Gouveia Dal Pino (2004). The mean values for the reference jet
simulation of Anderson et al. (2005) are higher, with  $j_j \simeq
450 \AU \km \s^{-1}$, and $e_j \simeq 2 \times 10^5 \km^2
\s^{-2}$. Using Anderson's et al. values will strengthen the
conclusions below. Let most of the jet material originate from an
annulus in the disk having a radius of $r_d$, and let the mass of
the central star be $M=1 M_\odot$. The specific angular momentum
and kinetic energy of the disk material at this radius are $j_d=30
(r_d/1\AU)^{1/2} \AU \km \s^{-1}$, and $e_d=450 (r_d/1\AU)^{-1}
\km^2 \s^{-2}$, respectively. By the assumptions of the
magneto-centrifugal models, a jets' angular momentum and energy
originate from the disk region where magnetic field lines are
anchored. Therefore, the ratio of mass outflow (in both jets
together) to mass accretion rate is constraint by angular momentum
conservation to be
\begin{equation}
\frac{\dot M_{j}}{\dot M_{\rm acc}} \le \frac{j_d}{j_j}=0.1 \left(
\frac{j_j}{300 \AU \km \s^{-1}} \right)^{-1} \left(\frac{r_d}{1
\AU} \right)^{1/2}, \label{jetj}
\end{equation}
and by energy conservation to be
\begin{equation}
\frac{\dot M_{j}}{\dot M_{\rm acc}} \le \frac{e_d}{e_j}=0.01
\left( \frac{v_j}{300 \km \s^{-1}} \right)^{-2} \left(\frac{r_d}{1
\AU} \right)^{-1}. \label{jete}
\end{equation}
The maximum mass outflow rate, for the chosen parameters, is
obtained for $r_{d-{\rm max}} \simeq 0.2 \AU = 45 R_\odot$, and
its value is $\dot M_{j-{\rm max}} = 0.046 \dot M_{\rm acc}$.
These values can be written in a general form as
\begin{equation}
r_{d-{\rm max}}=0.215 \left( \frac{v_j}{300 \km \s^{-1}}
\right)^{-4/3} \left( \frac{j_j}{300 \AU \km \s^{-1}}
\right)^{2/3} \AU, \label{diskr}
\end{equation}
and
\begin{equation}
\frac{\dot M_{j-{\rm max}}}{\dot M_{\rm acc}}=0.046 \left(
\frac{v_j}{300 \km \s^{-1}} \right)^{-2/3} \left( \frac{j_j}{300
\AU \km \s^{-1}} \right)^{-2/3}. \label{diskm}
\end{equation}

The calculation above shows that if the radial jet velocity
profiles observed by B2002 and C2004 are interpreted as being due
to jet rotation, then the required angular momentum and velocity
(or energy) of the jets (Cerqueira \& de Gouveia Dal Pino 2004),
constrain the mass outflow rate to be $< 5 \%$ of the accretion
rate. For a mass outflow rate of $\dot M_j \ga 0.05 M_{\rm acc}$,
as in RW Aur (Woitas et al.\ 2002, 2005) the disk loses its
Keplerian structure at very large radii, $\ge 0.2 \AU$. If the
disk maintain its Keplerian structure down to the central star, at
$r_d \simeq 0.01 \AU$, then only $\sim 1 \%$ of the accreted mass
can be blown in the jets; this is much below observed values. I
conclude that the required angular momentum for the rotating jet
model used by these authors is too high.

This conclusion is strengthened by the recent paper by Woitas et
al.\ (2005; this paper has become public after the first version
of the present paper has become public). In that paper Woitas et
al. study in greater detail the jets form RW Aur. The mass loss
rate in the two jets combined is $\dot M_j \sim 0.1 M_{\rm acc}$.
About half of the mass is lost from a wide annular region in the
disk, $0.4 \AU \la r_d \la 1.6 \AU$. This is compatible with the
reference model of Anderson et al. (2005), where half of the mass
is lost between the outer jet's foot point, $r_{d-{\rm max}}$ and
$0.25 r_{d-{\rm max}}$. Woitas et al. (2005) find that the jets
carry almost all the angular momentum lost by the disk material as
it flows inward. Woitas et al. (2005) don't give the value of the
energy carried by the jets in their model of RW Aur. Using their
supplied data, I estimate that the two jets in their model carry
almost all, or even more than, the energy lost by the disk
material as it flows from $r_d=1.6 \AU$ to $r_d=0.44 \AU$. This
implies that there is almost no dissipation in the process of
converting inflow energy to outflow kinetic energy in the outer
disk regions. This seems to be a too high efficiency for a process
involving dynamical strong magnetic fields. Garcia et al.\ (2001),
for example, find that the jets in their MHD launching model carry
$\la 60 \%$ of the energy released by the accreted mass in the
disk.

\begin{figure}
\includegraphics[width =135mm]{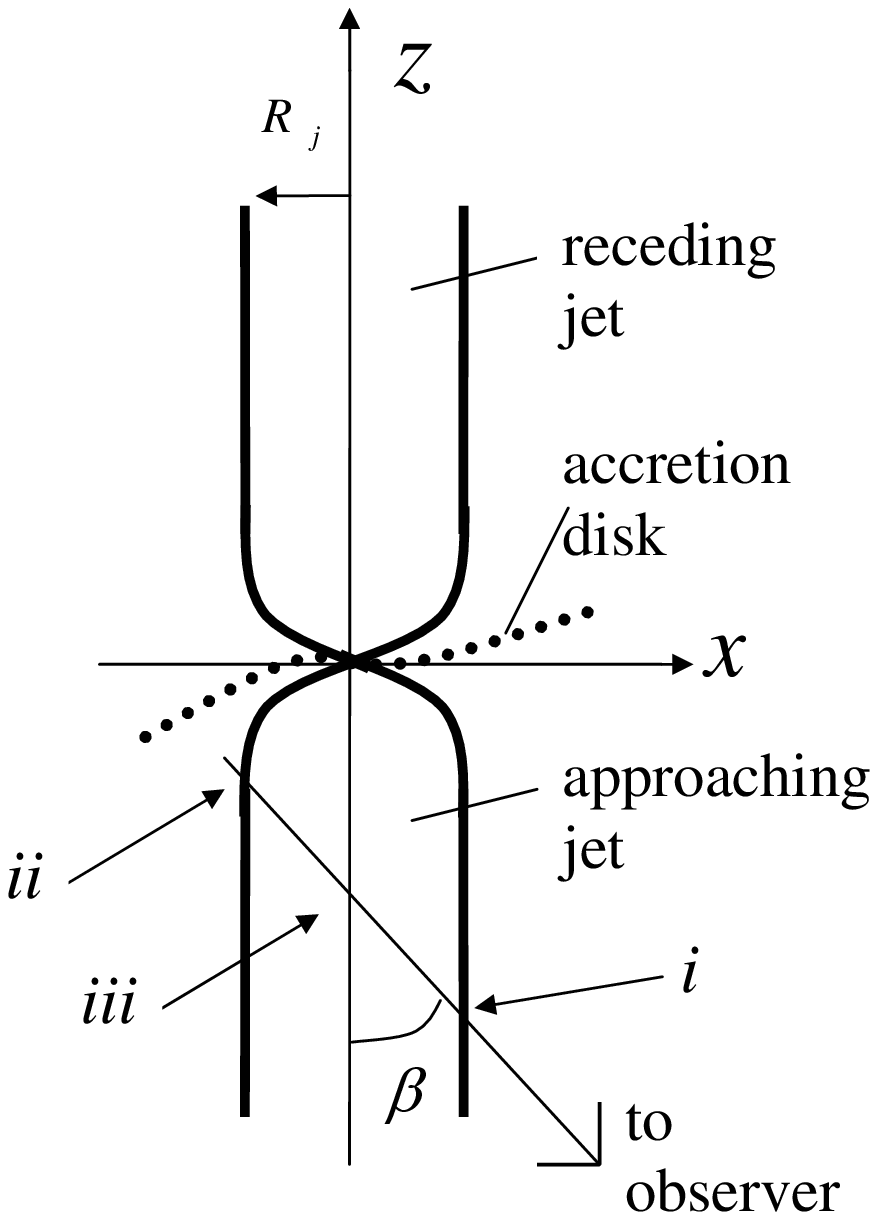}
\vskip 0.2 cm \caption{Schematic
drawing of the jet-disk-observer geometry in the plane defined by
the jets' symmetry ($z$) axis, and the line of sight. The solid
thick lines mark the boundary of the two jets. The dotted thick
line shows a cut through the twisted-tilted accretion disk. The
three points which are marked $i$, $ii$, and $iii$, are the three
points along each line of sight, at distance $y$ from the jet's
axis, for which the line of sight velocity is plotted in figure 3.
}
\end{figure}
\begin{figure}
\includegraphics[width =135mm]{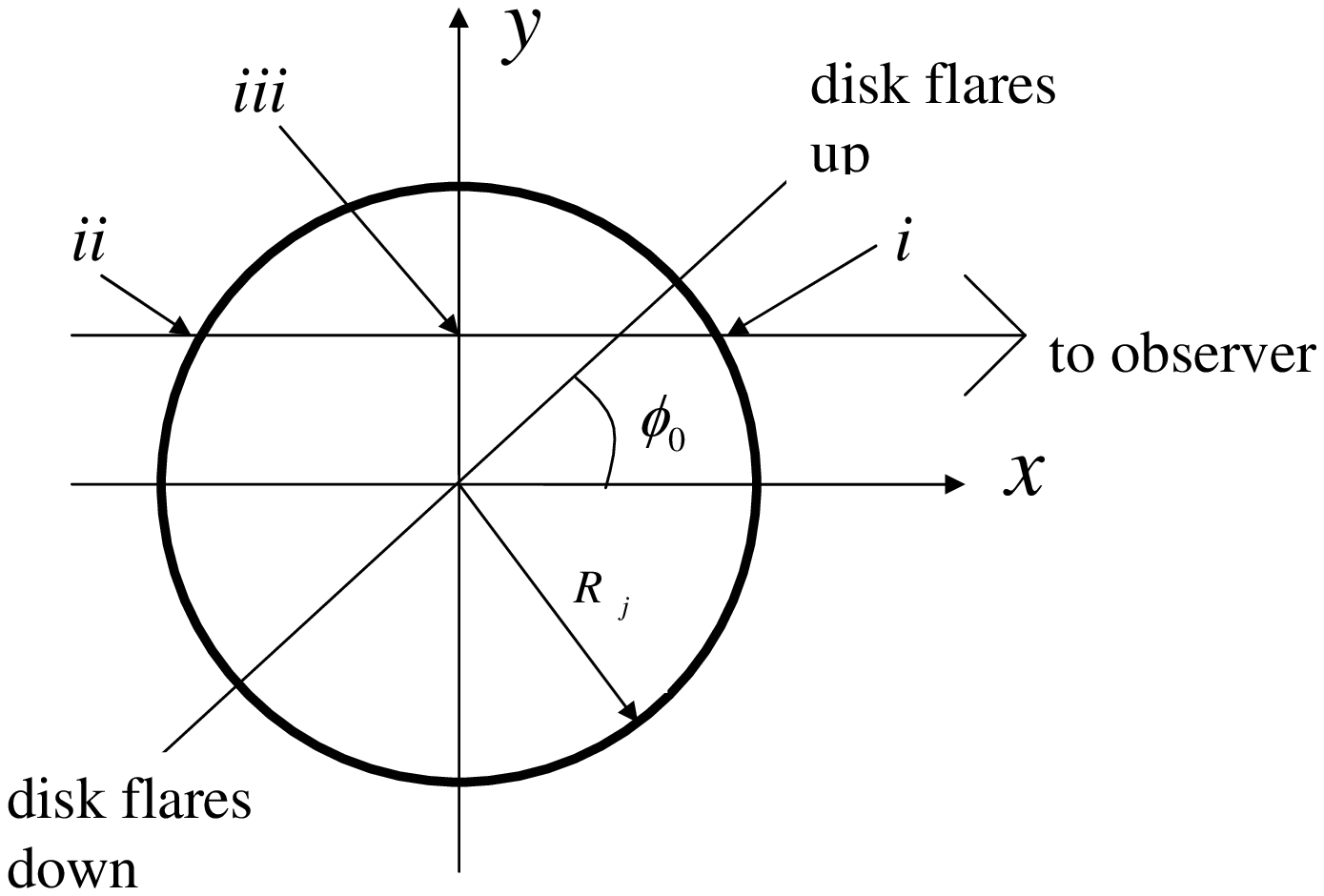}
\caption{Like Figure 1, but showing the plane perpendicular to the
jets' axis, and through the source of the jets.
The tilting and twisting of the accretion disk is monotonic,
with maximum flaring in the $+z$ direction at an angle $\phi_0$
to the $+x$ axis, and the maximum flaring in the $-z$ direction
occurring at an angle $\pi+\phi_0$ to the $+x$ axis.
}
\end{figure}
\begin{figure}

\includegraphics[width =135mm]{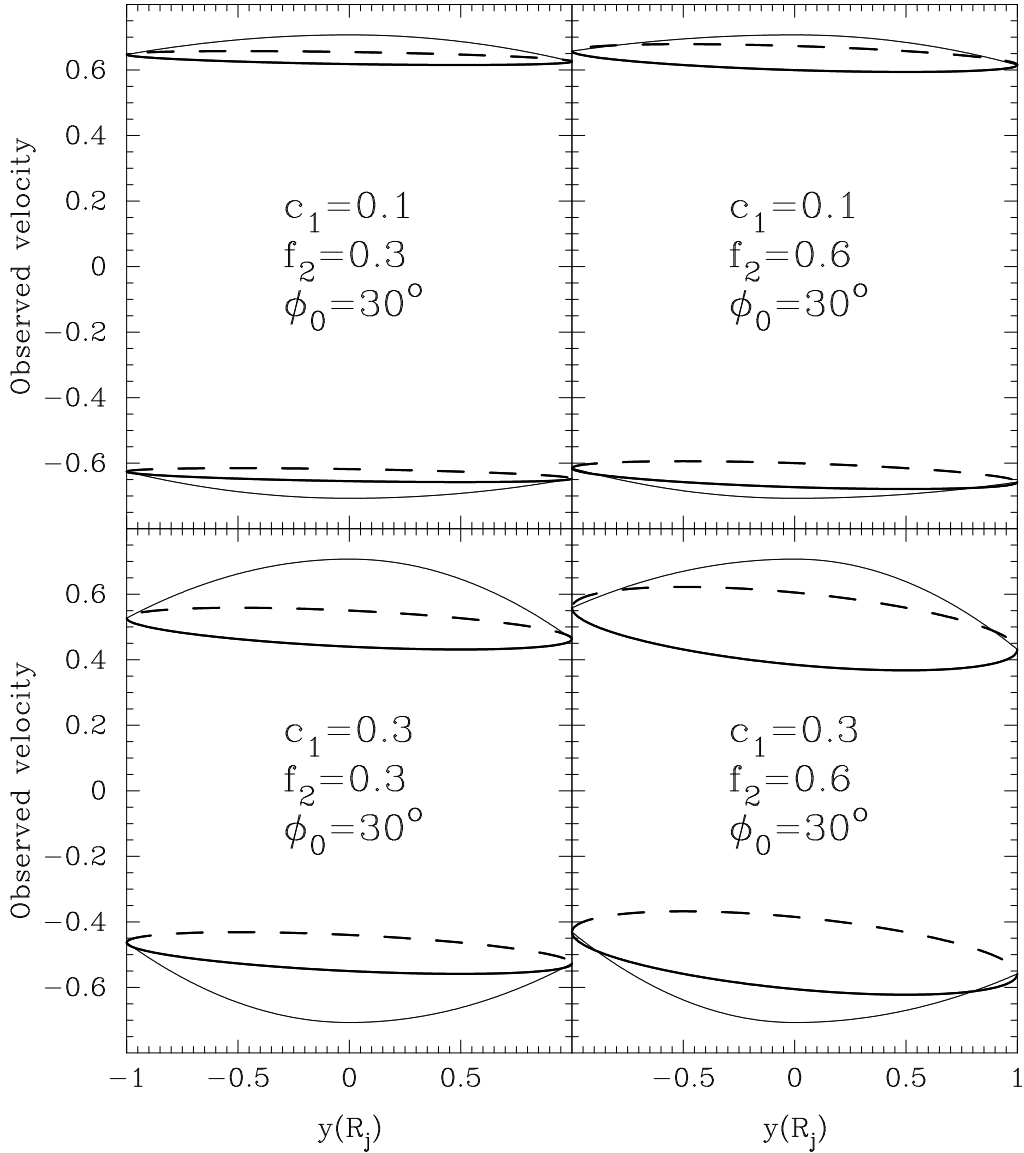}
\caption{The observed line of sight velocity across the jet, in units of
the real velocity along the jet's axis.
$y$ is the distance from the jet's symmetry axis, perpendicular to the line
of sight.
In each panel the upper 3 lines are for the receding jet, while the lower
3 lines are for the approaching jet.
Thick solid line is the velocity of the closest point in the jet along
the line of sight (eq. {\ref{vjeti}}; point $i$ in figs. 1 and 2);
Thick dashed line is the velocity of the farthest point in the jet
along the line of sight (eq. {\ref{vjeti}} with minus sign in
front of the square root; point $ii$ in figs. 1 and 2);
Thin solid line is the velocity of the jet's element closest to the
jet's axis, namely, on a line from the jet's axis and perpendicular
to the line of sight (eq. {\ref{vjetiii}}; point $iii$ in fig1. 1 and 2).
In all panels the jet axis in the $-z$ direction has an angle of
$\beta=45 ^\circ$ to the line of sight.
$\phi_0$ is defined in figure 2, while the meaning of $n$, $C_1$ and
$f_2$ can be seen in equation ({\ref{vjet1}});
$n=2$ in all cases drawn.
}
\end{figure}

\end{document}